\documentstyle[12pt]{article}
\font\titolo=cmbx10 scaled\magstep2
\font\titolino=cmbx12
\font\tsnorm=cmr12

\font\tscorsp=cmti10
\def\beginack
{\bigskip
\leftline{\titolino Acknowledgments.}
\nobreak\noindent}

\def\NPB{{\em Nucl. Phys. }}
\def\PLB{{\em Phys. Lett. }}  
\def\PRL{{\em Phys. Rev. Lett. }}
\def\PRD{{\em Phys. Rev. }}

\def\PRP{{\em Phys. Rep. }}

\def\z{Z\kern -4.6pt Z}
\def\dx{\int d^4x\ \sqrt{-g}\ }
\def\to{\rightarrow}
\def\lg{\left\{}
\def\lq{\left[}
\def\lt{\left(}
\def\rg{\right\}}
\def\rq{\right]}
\def\rt{\right)}
\def\lra{\leftrightarrow}
\def\ha{{1\over 2}}

\def\t{\tau}

\def\m{\mu}
\def\na{\nabla}
\def\n{\nu}

\def\pat{\partial^{2}_{t}}
\def\pax{\partial^{2}_{x}}
\def\P{\Phi}
\def\gmn{g_{\m\n}}
\def\ds{ds^{2}=}
\def\dx{\int d^2x\ \sqrt{-g}\ }

\def\ra{\rightarrow}
\def\l{\lambda}
\def\la{\l^2}

\def\be{\begin{equation}}
\def\ee{\end{equation}}
\def\bea{\begin{eqnarray}}
\def\eea{\end{eqnarray}}
\def\bc{\begin{displaymath}}
\def\ec{\end{displaymath}}
\def\lb{\label}

\begin{document}
\pagestyle{empty}
\null
\vskip 5truemm
\begin{flushright}
INFNCA-TH9802\\
March 1998 
\end{flushright}
\vskip 15truemm
\begin{center}
{\titolo 2D EXTREMAL BLACK HOLES AS SOLITONS}
\end{center}
\vskip 15truemm
\begin{center}
{\tsnorm Mariano Cadoni}
\end{center}
\begin{center}
{\tscorsp Dipartimento di Scienze Fisiche,  
Universit\`a  di Cagliari,}
\end{center}
\begin{center}
{\tscorsp Via Ospedale 72, I-09100 Cagliari, Italy,}
\end{center}
\begin{center}
{\tscorsp and}
\end{center}
\begin{center}
{\tscorsp  INFN, Sezione di Cagliari.}
\end{center}
\vskip 20truemm
\vfill
\begin{abstract}
\noindent
We discuss the relationship between   
two-dimensional (2D) dilaton gravity models  and sine-Gordon-like 
field 
theories. We show that there is a one-to-one correspondence between 
the solutions of 2D dilaton gravity and the solutions of a 
(two fields) generalization of the sine-Gordon model. In particular, 
we find a connection between the soliton solutions of the generalized 
sine-Gordon model 
and extremal black hole solutions of 2D dilaton gravity. 
As a by-product of our calculations we find a easy way to generate
cosmological solutions of 2D dilaton gravity.  
\begin{flushleft}
{PACS: 04.50 +h, 04.70.Bw, 11.10.Kk; \hfill}\\
\end{flushleft}
\end{abstract}
\vfill
\hrule
\begin{flushleft}
{E-Mail: CADONI@CA.INFN.IT\hfill}
\end{flushleft}
\eject
\pagenumbering{arabic}
\pagestyle{plain}
\section{Introduction}
\paragraph{}

The connection between black holes and 
non-perturbative structures of string theory, such as BPS 
solitons or  D-branes, has been one of the main ingredients of the 
last
exiting developments in string  theory \cite {RD,ST}. Black hole 
thermodynamics seems to have a natural explanation in terms of 
microscopic string and membrane physics \cite{ST}, opening new ways 
to address old (and new) fundamental problems of black hole physics.

On the other hand, the same fundamental problems of black hole physics 
have been  analyzed in the recent literature using low-dimensional 
gravity  models. In particular, 2D dilaton 
gravity models have been used to tackle challenging questions such as 
the ultimate fate of black holes or the loss of quantum coherence in 
the black hole evaporation process \cite{ST1}. Although no definitive 
answers to the above-mentioned problems have been found, 2D dilaton 
gravity 
models still provide an useful and simple framework to describe the 
4D black hole physics.

If one wants to use
the new ideas of string theory in the context of 2D dilaton gravity 
models, one has to investigate the role that solutions such as 
solitons 
play in these models. Moreover, for particular 2D dilaton gravity 
models 
we have  a direct relationship between BPS solitons of the 4D string 
effective theory and solutions of  the 2D model. For instance, the 
Jackiw-Teitelboim (JT) model \cite{JT} can be used to describe the   
S-wave 
sector of 
the extremal $D=4$, supersymmetric,  black hole solutions of models  
with 
dilaton coupling  $a=1/\sqrt 3$ \cite{MC,CM}. 

In a recent paper \cite{GK}, using the well-known correspondence   
between solutions  of the sine-Gordon theory and constant curvature 
metrics, Gegenberg and Kunstatter found 
  a relationship 
between black holes of JT dilaton gravity and solitons of the 
sine-Gordon field theory \cite {BC}. In this paper we explore the 
possibility to 
generalize this correspondence to generic 2D dilaton gravity models, 
whose solutions, in general, are not spacetimes of constant curvature.
We find that the field equations for 2D dilaton gravity are          
equivalent 
to those derived from a (two fields)  generalization of the           
sine-Gordon 
model.
From this correspondence we derive a connection between solitons of 
the generalized sine-Gordon model and  {\em extremal} black hole
solutions of 2D dilaton gravity. We also explain why  in the 
JT model this correspondence holds for the generic black hole 
solutions and not only for the extremal one. 

The structure of the paper is the following. In sect. 2 we show that 
the field equations of 2D dilaton gravity can be reduced to  those of
a generalized sine-Gordon model. In sect. 3 we derive the static 
solutions of the generalized sine-Gordon model. The conditions that 
have to be satisfied for these solutions to describe solitons, are 
also presented and implemented. In sect. 4 we discuss the 
relationship between the solitons and the black holes of the 2D 
dilaton gravity theory. In sect. 5 we use topological arguments to
classify the  soliton  solutions of the generalized 
sine-Gordon model. In sect. 6 we apply the general formulae that we 
have derived to some particular 2D dilaton 
gravity models. 
In sect. 7 we discuss  a by-product of our calculations,
namely 
a easy way to generate cosmological solutions of 2D dilaton gravity 
models.
Finally, in sect. 8 we present our conclusions.

\section{ 2D dilaton gravity and generalized sine-Gordon field        
theory}
\paragraph{}

Let us  consider the generic  two-dimensional dilaton gravity model.
Using  a Weyl rescaling of the metric and a reparametrization of the 
dilaton, one can write the most 
general action for the model in the form \cite{BM}:
\be\lb{e1}
S[\gmn ,\Phi] ={1\over 2\pi}\dx\left(\Phi R+
\l^2 V(\Phi) \right),
\ee
where $R$ is the curvature of the 2D spacetime, $V(\Phi)$ is an 
arbitrary function of the dilaton $\Phi$ and the 2D metric $\gmn$
has signature $(-1,1)$.
The field equations derived from the action (\ref{e1}) have the 
simple form \cite{BM},
\be\lb{e2}
R=-\l^2 {dV\over d\Phi},
\ee
\be\lb{e3}
\na_\m\na_\n\Phi-{\l^2\over 2}\gmn V=0.
\ee
By means of a suitable parametrization of the 2D metric one can       
map the 
solutions of the equations (\ref{e2}), (\ref{e3}) into the solutions 
of a generalized sine-Gordon model in 2D {\em Euclidean} space.
In fact, using the invariance of the theory under coordinate 
transformations we can write the 2D spacetime metric in the form, 
\be\lb{e4}
\ds -\sin^{2} \lt{u\over 2}\rt dt^{2}+\cos^{2} \lt{u\over 2}\rt      
dx^{2}, 
\ee
where $u=u(x,t)$. Taking into account that the corresponding      
curvature 
tensor is
\be
R=-{2\over \sin u}\lt\pat u +\pax u \rt,
\ee
one can show that the field equations (\ref{e2}), (\ref{e3}) are  
equivalent to the following set of equations in 2D Euclidean space,
\be\lb{e5}
\lt\pat u +\pax u \rt={\la\over 2}{dV\over d\P}\sin u,
\ee
\be\lb{e6}
\lt\pat \P +\pax \P \rt={\la\over 2}V\cos u.
\ee
Notice that eq. (\ref{e2}) is equivalent to eq. (\ref{e5}),          
one of the 
three equations in (\ref{e3}) translates into 
eq. (\ref{e6}), whereas the other two equations in (\ref{e3}) are 
integrability conditions for  the system (\ref{e5}), (\ref{e6}).

Instead of considering 2D dilaton gravity in Minkowski space, one can 
also start from the Euclidean formulation of this theory. 
In this case, one can easily demonstrate the equivalence of the {\em 
Euclidean}  field equations (\ref {e2}),(\ref{e3})  with the 
{\em Minkowskian} counterpart of eqs. (\ref{e5}), (\ref{e6}). The 
corresponding equations are obtained  performing the Wick rotation
$t\ra it$.

The field theory defined by the field equations (\ref{e5}),(\ref{e6}) 
can be considered as a (two fields) generalization of the sine-Gordon 
model. Eq. (\ref {e5}) reduces to the sine-Gordon equation for 
$V=\P$ \cite{GK} or, more generally, for constant configurations 
$\P_{0}$ of 
the dilaton, with  $V(\P_{0})=0,\, \lt dV/ d\P\rt (\P_{0})> 0$.
The field equations (\ref{e5}),(\ref{e6}) can be also obtained       
extremizing 
an action, which in 
Minkowski space has the form
\be\lb{e7}
S=\ha \int d^2 x \lt \P\Box u-{\la\over 2} V\sin u \rt,
\ee
where $\Box =-\pat +\pax$.

An unpleasant feature of the model (\ref{e7}) is that it describes a 
system of two scalar fields of opposite signature.
This can be easily seen performing a field redefinition that 
diagonalizes the kinetic energy of the fields,
\be\nonumber
\P= w+\phi,\qquad u=w-\phi.
\ee
Up to surface terms, the action (\ref{e7}) becomes,  
\be
S=-\ha \int d^2 x \lt \eta^{\m\n}\partial_\m w \partial_\n w -
\eta^{\m\n}\partial_\m \phi \partial_\n \phi +{\la\over 2}           
\sin (w-\phi) 
V(w+\phi)\rt,
\ee
where $\eta^{\m\n}=(-1,1)$.
The scalar field $\phi$ has negative  kinetic energy.

Let us conclude this section with some remarks on the correspondence
that we have established between the dilaton gravity model 
(\ref{e1}) and
the generalized sine-Gordon field theory (\ref{e7}).
Although there is a one-to-one correspondence (up to spacetime 
diffeomorphisms of the dilaton gravity theory) between the solutions 
of the
two theories, this does not mean that one can construct 
a metric solution that covers the whole 2D spacetime, once a 
solution of the 
eqs. (\ref{e5}),(\ref{e6}) is known.
In general, this is not possible owing to the particular 
parametrization
of the metric, given by eq. (\ref{e4}), which allows the metric 
coefficients 
$g_{tt}$ and
$g_{xx}$ to take values only in $]0,1]$.
However, one can  take analytic continuations of the solutions.
To this end, we can consider a parametrization of the metric 
obtained by 
replacing in eq. (\ref{e4}) the trigonometric with the hyperbolic 
functions,
\be\lb{e8}
\ds -\sinh^{2} \lt{u\over 2}\rt dt^{2}+\cosh^{2} \lt{u\over 2}
\rt dx^{2}. 
\ee
Starting from this expression for the metric, 
we can repeat the steps that led to 
the field equations (\ref{e5}),(\ref{e6}) and to the action   
(\ref{e7}).
What  we find now is an equivalence between the {\em Minkowskian 
(Euclidean)} 
dilaton gravity  field equations (\ref{e2}),(\ref{e3}) and the 
generalized
{\em Minkowskian (Euclidean)} sinh-Gordon field theory obtained
 by replacing 
in eqs. (\ref{e5}),(\ref{e6}) and  (\ref{e7}) the trigonometric 
 with the 
corresponding hyperbolic functions.
In conclusion, in order to have a complete  correspondence between  
2D spacetime structures of the dilaton gravity theory and    
solutions of 
sine-Gordon-like field theories, we need both the sine- and 
sinh-Gordon model.
 
\section{ Soliton solutions}
\paragraph{}
It is well known that the dilaton gravity model (\ref{e1}) admits     
solutions  that can  be interpreted as 2D black holes 
\cite{BH,BH1,CM}.
On the other hand, one 
expects the generalized sine-Gordon theory (\ref{e7}) to have         
soliton solutions that, in view of the results of the previous 
section, should be related to 2D black hole solutions. 
For arbitrary potential $V$ the existence of  soliton solutions is 
not a priori evident. We will therefore begin our discussion by 
answering  the question about the existence of solitons in the model
(\ref{e7}).

Solitons are non-singular field configurations that describe          
localized 
states of finite energy. Usually, necessary conditions for the 
existence of solitons can be found requiring the energy of the 
solution to be finite. Differently from the usual sine-Gordon model, 
in the case under consideration the energy functional is not positive 
definite. This is a consequence of the presence of a 
scalar field with negative kinetic energy.  From the action          
(\ref{e7}) 
it follows for the energy functional,
\be\lb{f1}
E(u,\P)=\ha \int^{\infty}_{-\infty}dx \lt \partial_t \P\partial_t u+ 
\partial_x \P\partial_x u + {\la\over 2}  V\sin u\rt.
\ee

Let us focus on static solutions of the field equations. 
To single out soliton solutions we require $E\ge 0$ and, at 
$x\to \pm \infty$,
\be\lb{f2}
\partial_{x}u=0, \qquad \partial_{x}\P=0.   
\ee
Note that for static solutions, the field equations imply 
$\partial_{x} u \propto V$ and $\partial_{x} \P \propto \sin u$
(see later eq. (\ref{f3})). Hence, conditions (\ref{f2}) imply       
also $V\sin u=0$ at
$x\to \pm \infty$.   
Conditions (\ref{f2})  are sufficient but not necessary for the 
finiteness of the energy. In fact, one can easily construct field 
configurations of finite energy for which $\partial_{x} u$ and/or 
$\partial_{x} \P$ are different from zero asymptotically. 
The existence of these configurations is related to the fact that the 
energy is not positive definite. It is not clear to us whether a 
soliton interpretation of these solutions is also possible.

The system of differential equations (\ref{e5}), (\ref{e6})          
admits the 
constant-field solutions $u=n \pi, \P=\P_{0}$, with $V(\P_{0})=0$.
One would like to identify some of these constant solutions as  
vacua of the theory. However, in the model under consideration one 
cannot define the vacuum in the usual way, just by looking for local 
minima of the potential energy. The problem is that our model        
contains 
a field with the wrong sign of the kinetic energy term. This fact     
makes
the usual  arguments about stability meaningless. On the other hand, 
because we are looking for soliton solutions, which tend 
asymptotically to some constant field configuration, we need to use a 
notion of vacuum of the model. The vacua of the model are defined as 
the zero energy, constant-field configurations $\P_{0}, u_{0}$ that 
satisfy, additionally, the following conditions. For $\P=\P_{0}$
the field  equations (\ref{e5}), (\ref{e6}) reduce to the usual      
sine-Gordon 
equation for $u$, whereas for $u=u_{0}$
they reduce to the equations of motion of a scalar field $\P$ with 
potential $V(\P)$.
These conditions single out, as vacua of the  model, the following 
constant values of the fields,
\be\lb{f2a}
u=2n\pi,\,\, n=0,\pm1,\pm2,\dots
 \qquad \P=\P_{0},\quad V(\P_{0})=0,\quad {dV\over d \P}( \P_{0})>0.
\ee 

For static configurations  the field equations (\ref {e5}),          
(\ref {e6}) can be 
integrated exactly.   
The first integral is
\be\lb{f3} 
u'=\l a V, \qquad \P'= {\l\over 2a} \sin u,
\ee
where $'={d/ dx}$ and $a$ is an integration constant.
A further integration gives the final form of the solutions,
\be\lb{f4}
\l\lt x-x_{0}\rt=\pm \int {d\P\over \sqrt{(K-c)\lq 1-                
a^{2}(K- c)\rq}},
\ee
\be\lb{f5}
\sin {u\over 2}= \pm a\sqrt{K-c},
\ee
where $K=K(\P)=\int^{\P} d\tau V(\tau)$ and $c,x_{0}$ are integration 
constants.

Eqs. (\ref{f4}), (\ref{f5})  do not give full account of the  general 
static solution of the field equations. In fact we have two,
two-parameters, families of solutions that are not contained in  
(\ref{f4}), (\ref{f5}).
The first family is obtained by taking a constant $u$ field,
\be\lb{f5a}
\l\lt x-x_{0}\rt=\pm \int {d\P\over \sqrt{(-1)^{n}K-b}},             
\qquad u=n\pi.
\ee  
The second family of solutions corresponds to a constant dilaton
field. These solutions exist only if $V(\P)$ has at least one 
zero $\P=\P_{0}$.  For $(dV/d\P)(\P_{0})> 0$ the field equations 
reduce to those of the usual sine-Gordon model,
\be
\nonumber
u''={\la\over 2}{dV\over d\P}(\P_{0}) \sin u.
\ee

Using eqs. (\ref{f1}) and (\ref{f3}),  one can calculate 
the energy of the solutions (\ref{f4}), (\ref{f5}). 
A straightforward calculation gives,
\be\lb{f6}
E= \l \lg K\lq\P(\infty)\rq-K\lq\P(-\infty)\rq\rg.
\ee

Having found an explicit form of the solutions, conditions (\ref{f2})
translate into restrictions on the admissible form of the dilaton    
potential
$V$ and on the values of the integration constants parametrizing 
the general solution. Using eq. (\ref{f3}) into eq. (\ref{f2}),
one gets 
\be\lb{f8}
V\lq\P(\pm\infty)\rq=0.
\ee
From equation (\ref{f4}) it follows (from now on we will set
 the physically
irrelevant integration constant $a= 1$) 
\be\lb{f7}
 \P'= \l\sqrt{\lt K-c\rt\lt 1-K+c\rt}.
\ee
Inserting  eq. (\ref{f7}) into eq. (\ref{f2}),
one finds the value of the integration constant $c$ for which
the solutions (\ref{f4}), (\ref{f5}) describe solitons,
\be \lb {f7a}
c=K\lq\P(\pm\infty)\rq.
\ee

It follows that the model (\ref{e7}) admits static soliton 
solutions, approaching for $x\to \pm \infty$  the constant
field configurations (\ref{f2a}) with $n=\pm 1$, if the equation 
$V(\P)=0$ admits at least one solution $\P_{0}=\P(\pm\infty)$ with 
$(dV/d\P) (\P_{0})>0$. The soliton  solutions 
are given by eqs. (\ref{f4}), (\ref{f5}) with $c=K(\P_{0})$.

Note that eqs. (\ref{f2}) are solved also by  $c=K[\P(\pm\infty)]    
-1$.
 However, it is evident from eq. (\ref{f5}) that the corresponding   
solutions 
tend asymptotically to $u=\pm \pi$. They can not be taken into 
consideration  if one requires, as we do here, 
that the soliton solutions approach asymptotically to one of the vacua 
(\ref{f2a}).

The energy of the soliton, given by equation 
(\ref {f6}), is zero if $\P(\infty)=\P(-\infty)=\P_0$, whereas it    
is different 
from zero if the potential $V$ has more then one zero and if 
$K[\P(\infty)]\neq K[\P(-\infty)]$.

\section{Soliton solutions and black holes } 
\paragraph{}

In the previous section we have derived soliton solutions of the 
model (\ref{e7}). The purpose of this section is to discuss the 
relationship between these soliton solutions and the black hole 
solutions of the 2D dilaton gravity model (\ref{e1}).

The solutions (\ref{f4}),(\ref{f5}) can be used  in a                
straightforward way 
to generate static 
solutions  
of the dilaton gravity theory. We just need to insert  
equations  
(\ref{f4}),(\ref{f5}) into the expression (\ref{e4}) for the metric      
of the 
2D spacetime.
Defining a new spacelike coordinate $r$,
\be\lb{f10}
{dr\over dx}=\sqrt{\lt 1-K+c\rt \lt K-c\rt},
\ee
we can write the spacetime metric and the dilaton in the form,
\be\lb{f11}
\ds -\lt K-c\rt dt^{2}+ \lt K-c\rt ^{-1} dr^{2},\quad \P=\l r.
\ee

This is the  general form for the static solution of 2D dilaton 
gravity 
\cite{BH1}. The    
parameter $c$ appearing  in eqs. (\ref{f11})  is related to the      
mass $M$ of
the solution, $c=2M/\l$. Under suitable conditions the solutions 
(\ref{f11}) can be interpreted as 2D black holes.
Hence,  every dilaton gravity model, whose dilaton potential  $V(\P)$ 
has at least one zero $\P=\P_{0}$ with $(dV/d\P) (\P_{0})>0$,
 has a solution, with mass given by
$M=\l K(\P_{0})/2$, that can be realized as  a soliton solution 
(\ref{f4}),(\ref{f5}).
Conversely, given a soliton solution of the generalized sine-Gordon
model (\ref{e7}), one can always construct a 
solution of the 2D dilaton gravity model (\ref{e1}), which has the
form (\ref{f11}) with $c=K(\P_{0})$.

If the 
solution of the 2D dilaton gravity model describes a black hole,     
the soliton 
solution can be put in correspondence with a 2D black hole.
Furthermore, one can easily show that the  solutions of 2D dilaton   
gravity model
 that 
can be realized as solitons of the model (\ref{e7}) 
describe spacetimes with no event horizons, which eventually can be
interpreted as 
 {\em extremal} black holes. In fact, assuming that the 
function $K(\P)$ is monotonic in the considered range of variation 
of $\P$ (this condition is necessary for the black hole 
interpretation) it follows that the function $F(\P)=K(\P) -K(\P_{0})$ 
 has 
only one zero, at $\P=\P_{0}$, which owing to condition (\ref{f8}) is 
actually a double zero, $(dF/d\P)(\P_{0})=V(\P_{0})=0$. As a         
consequence 
the Killing vector associated with the solution (\ref{f11}) with     
$c=K(\P_0)$
cannot become 
spacelike anywhere, i.e.  the spacetime has no event horizons.

The previous statement seems to contradict the results of ref.
\cite{GK} for the JT model. In ref. \cite{GK} it has been shown that 
every black solution of the JT theory  
can be put in correspondence with a soliton solution of the sine     
Gordon 
model. Later in this paper
we will discuss this point in detail and we will argue that this 
 behavior is due to a peculiar feature  of the black hole 
solutions of the JT model.

As we have already noted in sect. 2 the correspondence between black 
holes 
and soliton solutions is limited to a region of the 2D spacetime. 
Using  the coordinate $r$ defined in  eq. (\ref{f10}) to             
parameterize the 2D 
spacetime  and  taking into account eq. (\ref{e4}), one can 
easily realize that this  region is defined by  $ 0\le (K(\l r) 
-2M/\l)\le 1$. From the discussion in sect. 2 it is also evident that
the spacetime  region $(K(\l r) -2M/\l)\ge 1$ can be put in 
correspondence with  a generalized sinh-Gordon model. 

\section {Topological analysis}
The static configurations  (\ref{f4}),(\ref{f5}) describe            
time-independent 
 soliton solutions
of the model (\ref{e7}). One can generate 
propagating soliton solutions performing in eqs. (\ref{f4}),         
(\ref{f5})
the transformation
\be
x\to \gamma \lt x \pm v  t\rt,
\ee
where $\gamma =1/\sqrt {1-v^{2}}$ and  $v$ is the velocity of        
propagation 
of the soliton.    
Apart from the solutions (\ref{f4}), (\ref{f5}) there are no other 
time-independent soliton 
solutions  of 
the theory (\ref{e7}). Being described by non-linear equations the 
behavior of many-soliton system is always time-dependent.
The admissible number of solitons can be 
determined in the usual way,  using  topological properties of 
the symmetry group of 
the model (\ref{e7}). Conditions (\ref{f2}) imply that every 
soliton solution of the field equations tends asymptotically to      
one of 
vacuum configuration (\ref{f2a}). Therefore the number of 
admissible solitons is determined by the number of ways in which the 
points $x=\pm\infty$ (the zero-sphere) can be mapped into the 
manifold  of constant-field configurations (\ref{f2a}). We have, 
therefore, a one-to-one correspondence between solitons and elements 
of the homotopy group
$\pi_{0}(G/H)$, where $G$ and $H$ are 
respectively the group of internal symmetries of the model and the      
residual 
symmetry of the vacua (\ref{f2a}).
$G$ and  $H$ depend on the form of the dilaton 
potential $V$. For generic $V$ the model (\ref{e7}) has an invariance 
group $G=Z\times Z_{2}$, where $Z$ and $Z_{2}$ are, respectively, the 
infinite discrete group of  translations and the finite group of 
inversions of the field $u$:
\be\lb{g1}
u\to u+2\pi n, 
\qquad u\to -u.
\ee
Note that the effect of the transformation $Z_{2}$ on the action 
(\ref{e7}) is to flip its sign.
The vacua (\ref{f2a}) have a residual $Z_{2}$ symmetry, so that 
the homotopy group  is 
\be
\pi_{0}\lt{Z\times Z_{2}\over  Z_{2}}\rt=\pi_{0} (Z)=Z.
\ee
The same result holds for  the usual sine-Gordon model.
For particular choices of the dilaton potential $V$ we can have 
$ Z\times Z_{2}\subset G$ and $\pi(G/H)\neq Z$. In the next section 
we will give examples of this kind of  behavior.

It is well known that field theories that admit   soliton solutions   
have  
conserved currents, corresponding to conserved topological charges.
For the model under consideration we can define two independent 
topological currents, 
\be\lb{g2}
J_{(u)}^{\nu}=\epsilon^{\n\m}\partial _{\mu}u,\qquad 
J_{(\P)}^{\nu}=\epsilon^{\n\m}\partial _{\mu}\P.
\ee
The associated topological charges are
\bea\lb{g3} \nonumber
N_{(u)}&=&{1\over 2\pi}\int_{-\infty}^{\infty} dx J_{(u)}^{0}= 
{1\over 2\pi}\lq u(\infty)- u(-\infty)\rq,\\
N_{(\P)}&=&{1\over 2}\int_{-\infty}^{\infty} dx J_{(\P)}^{0}= 
{1\over 2}\lq \P(\infty)- \P(-\infty)\rq.
\eea

\section{Some examples}

In this section we apply the general formulation described in the    
previous 
sections to some particular 2D dilaton gravity models. We will 
consider three examples: the JT theory defined by a dilaton potential 
$V(\P)=2\P$, a model with a degree-three polynomial potential, 
$V(\P)=4\P(\P^{2}-1)$ and a $\sinh\P$-model with potential           
$V(\P)=\sinh 
2\P$.
\subsection {The JT model}
The JT model is characterized by the potential $ V(\P)= 2\P$, which
has a zero at $\P=0$ with $dV/d\P=2$. According to the discussion of 
the previous sections, we expect the corresponding generalized 
sine-Gordon model to have soliton solutions, described by eq. 
(\ref {f4}), (\ref{f5}) with $c=K(0)=0$. With $K=\P^{2}$ and $c=0$ 
we can perform 
explicitly the integration in eq. (\ref{f4}), the final form of the 
soliton solution is
\be\lb {h1}
u= \pm 4 \arctan \exp \l\lt x-x_{0}\rt, \qquad \P^{-1}=              
\cosh \l\lt x-x_{0}\rt.
\ee 

The previous solution has zero energy.  Depending on  the  sign 
in eqs. (\ref{h1}), we have solitons  and antisolitons.
The  corresponding solution of the JT model is the extremal          
black hole 
solution (the ground state of the model) with $M=0$,
\be\lb{h2}
\ds -\P^{2}dt^{2}+ \P^{-2}dr^{2},\qquad \P=\l r.
\ee
The symmetry group of the model is 
 $G=Z_{2}^{(\P)}\times Z_{2}^{(u)}\times Z^{(u)}$,
where $Z_{2}^{(\P)}$ is the transformation $\P\to -\P$ and 
$Z_{2}^{(u)}\times Z^{(u)}$ is the symmetry group of the sine-Gordon 
model given by eq. (\ref {g1}). The vacua $\P=0, u= 2n \pi$ have a 
residual symmetry  $Z_{2}^{(\P)}\times Z_{2}^{(u)}$, so that for the 
homotopy group we have,
\be\lb{h3}
\pi_{0}\lt{G\over H}\rt= Z.
\ee
The model admits soliton solutions with topological charges 
(\ref{g3}) given by $N_{(\P)}=0, N_{(u)}=\pm n, n=0,1 \dots$, where  
the 
plus sign refers to soliton and the minus sign to antisoliton 
solutions. In particular the time-independent solitons (\ref{h1})    
have 
topological charge   $ N_{(\P)}=0, N_{(u)}=\pm 1$, respectively for 
the soliton and the antisoliton.

That the soliton solutions (\ref{h1}) can be put in 
correspondence with  solutions of the JT theory, has previously been 
demonstrated  by Gegenberg and Kunstatter \cite {GK}. They have also 
shown that the soliton solution (\ref{h1}) can be put in             
correspondence with 
every ($M=0$ or $M>0$)  black hole solution of the JT theory, not 
only with the $M=0$ extremal solution (\ref{h2}). 
This seems in contradiction with our result of  
sect. 4, stating that only 
2D spacetimes with no event horizons are in correspondence with      
soliton
solutions of the theory (\ref{e7}). The two results are not in 
contradiction because  of a well-known feature of the JT model,      
namely 
the fact that all the static solutions of the 
JT theory are different parametrization of the same manifold,  
2D anti-de Sitter spacetime  (see for example 
\cite{CM} and references therein). There is always a coordinate 
transformation relating the $M=0$ solution (\ref{h2}) with the       
generic 
$M>0$ black hole solution. Using these coordinate transformations one 
can always map the soliton (\ref{h1}) into a generic static solution  
of the
JT model.   This feature is a peculiarity of the JT 
model that is not expected to hold for a generic 2D dilaton gravity  
model.

\subsection {Degree-three polynomial potential}

Let us now consider a model with $V=4\P\lt \P^{2}-1\rt$. The model   
has the 
same symmetry group as the JT model, namely 
$G=Z_{2}^{(\P)}\times Z_{2}^{(u)}\times Z^{(u)}$. Differently from 
the JT case, now the vacua $\P=\pm 1, u= 2n\pi $ break the           
$Z_{2}^{(\P)}$ 
symmetry, the residual symmetry group being now 
 $H=Z_{2}^{(u)}$. Therefore, the 
model will admit  a richer spectrum of soliton solutions. The        
homotopy 
group is $\pi_{0}(G/H)= Z^{2}\times Z$, so that in this case we can  
have a  different from zero topological charge $N_{\P}$. The soliton 
solutions can be classified using a 
topological charge vector 
$X=(N_{\P}, N_{u})$ with entries   $N_{\P}, N_{u}$ given by eq 
(\ref{g3}). The topologically trivial solutions, characterized by
$X=(0,0)$, are given by 
the vacua $\P=\pm 1, u=2n\pi$. The soliton solutions labeled by  
$X=(0,\pm n)$ are usual sine-Gordon solitons, i.e they  have a       
constant,
$\P=\pm 1$, dilaton, whereas $u$ is the solution of the sine-Gordon 
equation 
\be
-\pat u+\pax u=4\la  \sin u.
\ee  
The solitons characterized by $X=(\pm 1,0)$ are {\em kinks}.         
They have 
a constant $ u$ field, $u=2n\pi$, whereas $\P$ is solution of        
the equation
\be
-\pat \P+\pax \P=2\la \P(\P^{2}-1),
\ee
which admits the {\em kink } solution 
\be
\P= \pm \tanh \l(x-x_{0}).
\ee

Finally, we have the soliton solutions with $X=(\pm 1, \pm n)$.
The solitons with $n=1$ are given by eqs. (\ref {f4}), (\ref{f5})    
with 
$K= \P^{4}-2\P^{2}$ and $c=K(\pm 1)= -1$.  With these positions 
equation (\ref{f4}), (\ref{f5}) can be easily integrated. Here,      
we do not write 
down the resulting  expression 
because it is  rather  cumbersome and not essential for our purposes.
This soliton solution has zero energy and  is in correspondence with   
 an 
extremal  black hole solution
of the 2D dilaton gravity model,
\be\lb{h4}
\ds -\lt \P^{2}-1\rt^{2}dt^{2}+\lt \P^{2}-1\rt^{-2}dr^{2}, \quad     
\P=\l 
r.
\ee
The previous metric is extremal  because the 
 generic  solution of the 2D dilaton gravity model
has an horizon for $\P^{2}=1+\sqrt{1+ c}$.
Solution (\ref{h4}) is the $c=-1$ solution and it is, therefore,
the extremal one.  

\subsection { The $\sinh \P$-model}
As a last example,  we consider a model with $V(\P) = \sinh 2\P$.
The vacua, the symmetry group  $G$ and the group of 
residual symmetry are the same as those of the JT theory. As a 
consequence,  also the homotopy group is the same, $\pi_{0}(G/H)=Z$.
The soliton solutions with topological charge $N_{u}=\pm1$ can be 
easily obtained using eqs. (\ref{f4}), (\ref{f5})
with $K(\P)=(\cosh 2\P)/2$ and 
$c=K(0)=1/2$. These soliton solutions have zero energy and 
 can be written in the form,
\bea
{u\over 2}&=& {\pi \over 2} +\arctan \lq\sqrt 2\sinh \l \lt 
x-x_{0}\rt\rq,\\
\P&=& {\mathrm{Artanh}} \lq \sqrt2 \cosh \l\lt x-x_{0}\rt \rq^{-1}.
\eea
The corresponding black hole solution of the 2D dilaton gravity model 
 is
\be 
\ds -\sinh^{2} \P dt^{2}+ \sinh^{-2} \P dr^{2}, \quad \P=\l r.
\ee
Again, this is an extremal solution because  the horizon of the      
general 
solution is located at $\sinh \P =\sqrt{c-1/2}$.  
 
\section {Cosmological solutions of 2D dilaton gravity}

An interesting  by-product of our discussion 
 is an easy way to generate general cosmological  solutions 
of 2D dilaton gravity.  Although 
cosmological solutions have already been  found for particular 2D 
dilaton gravity models \cite{TV,CC},  till now the solution for the 
generic model has not been derived.

Cosmological solutions of the model (\ref{e1}) can be easily found 
using the parametrization of the 2D metric given by eq.(\ref{e4})    
and the 
equivalence between the field equations (\ref{e2}),(\ref{e3})
in Minkowski space and
the field equations  (\ref{e5}), (\ref{e6}) in Euclidean space. 
The point is that,  being 
the  equations (\ref{e5}),(\ref{e6}) written in Euclidean space, they 
maintain their form if we interchange $x\lra 
t$. Cosmological solutions of 2D dilaton gravity  can be, therefore,
 generated from the static solutions 
just by interchanging $x\lra t$. From eqs. (\ref{f4}), (\ref{f5})
it follows the 
general form of the cosmological solutions,
\bea\nonumber
 \ds &-&\sin^{2} \lt{u(t)\over 2}\rt dt^{2}+\cos^{2} \lt{u(t)        
\over 2}\rt 
 dx^{2},\\ \nonumber      
\sin \lt{u\over 2}\rt&=&\pm \sqrt{K-c},\\ \nonumber
\l\lt t-t_{0}\rt&=&\pm \int {d\P\over \sqrt{(K-c) \lq 1- (K- c)\rq}}.
\eea

Introducing the cosmological time $d\t=\sin(u/2) dt$, the solutions 
take the form
\bea\lb{h6}\nonumber
\ds &-& d\t^{2}+\cos^{2} \lt{u\over 2}\rt 
 dx^{2},\\ \nonumber      
\sin \lt{u\over 2}\rt&=&\pm \sqrt{K-c},\\
\l\lt \t-\t_{0}\rt&=&\pm \int {d\P\over \sqrt{ 1- (K- c)}}.
\eea

Let us give two  examples of the application of eqs. (\ref{h6}).
Consider first the cosmological solutions of the JT model.
With $K=\P^{2}$,  we can perform the integration in (\ref{h6}),      
we get
\bea\lb{h7}\nonumber
\ds &-& d\t^{2}+ A^{2}\sin^{2}\l (\t-\t_{0}) dx^{2},\\  
\P&=& A\cos \l (\t-\t_{0}),
\eea
where $A=\sqrt {1+c}$. This form of the solution has already been     
found 
using a different method in ref. \cite{CC}.

In the case of a model with a exponential  potential 
$V=\exp \P$, eqs. (\ref{h6}) give
\bea
\ds &-& d\t^{2}+ A^{2}\lq \tanh ^{2}{\l A\over 2} (\t-\t_{0})\rq     
dx^{2},\\  
\P&=& - 2\ln \cosh {\l A\over 2} (\t-\t_{0}) +const,
\eea
 where $A$ is an integration constant given as in eq. (\ref{h7}).
 
\section {Conclusions}
In this paper we have shown that one can use solutions of a          
generalized 
sine-Gordon model to describe the classical dynamics of 2D dilaton   
gravity.
As a consequence, 
we have found for a broad class of extremal 2D black holes   
an underlying solitonic solutions that, in principle, can be used to
describe their  classical behavior.
Our results seem to indicate that there is a deep connection between 
extremal configurations of 2D  black hole and solitonic states.
However, for 2D dilaton gravity the situation seems rather different 
from that one has  for  supergravity theories in  $D\ge 4$. In the   
latter case
the condition that the solutions preserve at least $N=1$             
supersymmetry 
implies that they are BPS states, i.e that they
saturate at least one Bogomol'nyi  bound. In this way the extremal   
black 
hole solutions have a natural interpretation as BPS solitons.
In the case of 2D dilaton gravity the absence of  bounds related 
with symmetries of 
the model  (such as the  Bogomol'nyi
bound) makes the notion of extremality model dependent.
The simplest way to obtain these bounds would be to consider a 
supersymmetric extension of our model. Previous investigations of 
the supersymmetric  sine-Gordon model indicate, however, that in     
this model 
the BPS states are not solitons \cite{RN}. It would be of interest to 
find out if this apply also to the supersymmetric extension of the 
model (\ref{e7}).

Apart from these difficulties, our approach represents just a first    
step 
in the analysis of the role that solitons play in 2D dilaton gravity.
It leaves many open question. We have seen  that there is 
a sine-Gordon-like theory underlying the classical behavior of 2D     
extremal 
black holes. However, the most interesting questions in the black    
hole 
physics appear at the semiclassical (or full quantum) level. The      
natural
development of our approach would be to use the 
solitons to describe the semiclassical and the quantum dynamics      
of  2D 
extremal black holes. For instance, one could try to  use soliton 
physics  to describe near extremal 2D black holes, to investigate the 
black hole evaporation process and to give a microscopic 
interpretation of the entropy of the hole.  
\smallskip

\beginack

I am grateful to G. Kunstatter  for interesting discussions.

\end{document}